\documentclass[preprint,aps,prb,showpacs]{revtex4}
\usepackage{bm}
\usepackage{epsfig}

\newcommand{\nix}[1]{}

\begin{document}

\title{ Zero-bias spin separation}
\author{Sergey D.~Ganichev$^{1}$, Vasily V.~Bel'kov$^{1,2}$, Sergey A.~Tarasenko$^2$, Sergey N.~Danilov$^1$,
Stephan~Giglberger$^1$, Christoph~Hoffmann$^1$, Eougenious L.~Ivchenko$^2$, Dieter~Weiss$^1$, Werner~Wegscheider$^1$
 Christian~Gerl$^{1}$, Dieter~Schuh$^{1}$, Joachim Stahl$^{1}$, Joan~De~Boeck$^3$, Gustaaf~Borghs$^3$, 
\& Wilhelm~Prettl$^1$ \\
}
\affiliation{$^1$Fakult\"{a}t Physik, University of Regensburg,
93040, Regensburg, Germany}
\affiliation{$^2$A.F.~Ioffe Physico-Technical Institute, Russian
Academy of Sciences, 194021 St.~Petersburg, Russia}
\affiliation{$^3$IMEC, Kapeldreef 75, B-3001 Leuven, Belgium}

\date{\today}


\maketitle

\newpage

{\bf Spin-orbit coupling provides a versatile tool to generate and
to manipulate the spin degree of freedom in low-dimensional
semiconductor structures. The spin Hall effect, where an
electrical current drives a transverse spin current and causes a
nonequilibrium spin accumulation observed near the sample 
boundary~\cite{Awschalom,Wunderlich}, the spin-galvanic effect, where a
nonequilibrium spin polarization drives an electric current~\cite{Ganichev1}, 
or the reverse process, in which an electrical current
generates a nonequilibrium spin polarization~\cite{Ganichev2,Silov,Awschalom3}, 
are all consequences of spin-orbit
coupling. In order to observe a spin Hall effect a bias driven
current is an essential prerequisite. The spin separation is
caused via spin-orbit coupling either by Mott scattering
(extrinsic spin Hall effect) or by
spin splitting of the band structure
(intrinsic spin Hall effect). Here we provide evidence for an elementary
effect causing spin separation which is fundamentally different from
that of the spin Hall effect.
In contrast to the spin Hall effect
it does not require an electric current to flow:
It is  spin separation  achieved by spin-dependent scattering of
electrons in  media with suitable symmetry.
We show that by  free carrier (Drude) absorption of terahertz
radiation spin separation is achieved in a wide range of
temperatures from liquid helium up to room temperature. Moreover
the experimental results give evidence that simple electron gas
heating by any means is already sufficient to yield spin
separation due to spin-dependent energy relaxation processes of
nonequilibrium carriers.}

Scattering of electrons involves a transition from a state
with wavevector $\bm{k}$ to a state with wavevector $\bm{k}^\prime$ which is usually
considered to be spin-independent. However, in gyrotropic media, e.g.  
GaAs quantum wells or heterojunctions, spin-orbit interaction
adds an asymmetric spin-dependent term  to the scattering probability~\cite{spinHall}. The asymmetric 
spin-dependent scattering matrix element is linear in
wavevector~\cite{footnote4}  $\bm{k}$ and the Pauli spin matrices $\bm{\sigma}$.
Microscopically this term is caused by
structural inversion asymmetry (SIA) and/or bulk inversion asymmetry (BIA).
While the asymmetry of electron scattering  can
cause spin currents to flow, it does not modify the energy
spectrum.

A process actuating spin separation is illustrated in Fig.~1~(a) and
involves  Drude absorption of radiation.  Drude absorption is caused
by indirect intraband optical transitions and includes a momentum
transfer from phonons or impurities  to electrons to satisfy momentum
conservation.
Figure~1~(a)
sketches the process of Drude absorption via virtual states for a
spin-up subband ($s=+1/2$, left panel) and a spin-down subband
($s=-1/2$, right panel) of a quantum well containing a two-dimensional
electron gas.

Vertical arrows indicate optical transitions from the initial
state with  electron wavevector $k_x = 0$ while the  horizontal
arrows describe an elastic scattering event to a final state with
either positive or negative electron wavevector $k^\prime_x$. While, for simplicity of
illustration, we have only shown transitions starting from
$k_x$~=~0, the arguments given here are valid for arbitrary $k_x$.
Due to the spin dependence of scattering, transitions to positive
and negative $k^\prime_x$-states occur with different
probabilities. This is indicated by horizontal arrows of different
thicknesses. Since the asymmetric part of electron scattering is
proportional to components of the vector product 
$[\bm{\sigma}\times\bm{k}^\prime]$  higher
and lower probabilities for scattering to positive or negative
$k^\prime_x$ get inverted for a spin-down subband compared to
that of a spin-up subband~\cite{footnote1}. Similarly, also relaxation of excited carriers 
is asymmetric as is sketched in Fig.~1~(b). Since this mechanism causes 
only a polarization independent background  signal 
in the experiments discussed below, the discussion will first be  focused on the mechanism 
displayed in Fig.~1~(a).

The asymmetry causes an imbalance in the distribution of
photoexcited carriers in the spin subbands ($s=\pm 1/2$) between
positive and negative $k^\prime_x$ states, which in turn yields
electron flows $\bm{i}_{\pm 1/2}$ within each spin subband~\cite{TI_jetplett}.
However, the charge currents,  $\bm{j}_+ = e\bm{i}_{1/2}$ and
$\bm{j}_- = e\bm{i}_{-1/2}$, where $e$ is the electron charge,
have opposite directions because $\bm{i}_{+1/2} = -\bm{i}_{-1/2}$ and therefore cancel each other.
Nevertheless, a spin current $\bm{J}_{\rm spin} =
\frac{1}{2}(\bm{i}_{+1/2} - \bm{i}_{-1/2})$ is generated since
electrons with spin-up and spin-down move in opposite directions.
This leads to a spatial spin separation and spin accumulation at the
edges of the sample.

By application of a magnetic field which polarizes spins, the spin
current gets detected as charge current. This is analogous to
spin-dependent scattering in transport experiments:  Mott
scattering of unpolarized electrons causes the extrinsic spin Hall
effect, whereas in a spin polarized electron gas a charge current,
the anomalous Hall effect, can be observed. In a spin polarized
system,
the two fluxes $\bm{i}_{\pm 1/2}$, which are proportional to the
spin-up and spin-down free carrier densities, $n_{\pm 1/2}$, cease
compensating each other and yield a net electric
current~\cite{footnote2}
\begin{equation}
    \label{current1}
\bm{j} = e (\bm{i}_{+1/2}  + \bm{i}_{-1/2}) = 4 e S \bm{J}_{\rm spin} \:,
\end{equation}
where $S = \frac{1}{2}(n_{+1/2} - n_{-1/2})/(n_{+1/2} + n_{-1/2})$
is the average spin. An external magnetic field $\bm B$
results in different
populations of the two spin subbands  due to the Zeeman effect. In equilibrium the average spin is
given by
\begin{equation}\label{spin}
{\bm S} = - \frac{g \mu_B \bm{B}}{4 \bar{\varepsilon}}\:.
\end{equation}
Here $g$ is the electron effective $g$-factor, $\mu_B$  the Bohr
magneton, $\bar{\varepsilon}$  the characteristic electron energy
being equal to the Fermi energy $\varepsilon_F$, or to the thermal
energy $k_B T$, for a degenerated and a non-degenerated
two-dimensional electron gas, respectively.

In order to demonstrate the existence of the spin 
current due to asymmetric scattering described above we carry out
the following experiment: Drude absorption is achieved using
linearly polarized terahertz radiation directed along the growth
direction of a (001)-oriented heterostructures. The equilibrium
spin polarization is obtained by an in-plane magnetic field
$\bm{B}$, which shifts the two parabolas of Fig.~1~(a) vertically
by $\pm g \mu _{B} B/2$. The photocurrent is measured both in the
directions perpendicular and parallel to the magnetic field. The
chosen experimental conditions exclude other effects known to
cause photocurrents: Since linearly polarized radiation is used,
all helicity dependent spin photocurrents, 
such as the spin-galvanic effect~\cite{Ganichev2} and 
the circular photogalvanic effect~\cite{Ganichev3}, are absent.
In addition, a possible photon drag effect and the linear
photogalvanic effect are forbidden by symmetry for normal
incidence on (001)-grown heterostructures~\cite{bookGP}.

The experiments are carried out on both, MBE-grown (001)-oriented
$n$-type GaAs/AlGaAs and InAs/AlGaSb two-dimensional structures.
The parameters of the investigated samples are given in Table~I.
Two pairs of
ohmic  contacts at the center of the sample edges and lying along
the $x
\parallel$ [1$\bar{1}$0] and $y \parallel$ [110]
directions have been prepared to measure the photocurrent (see
inset in Fig.~2).
A high power pulsed molecular terahertz laser has been used as 
radiation source delivering 100~ns pulses with  radiation power
$P$ up to 1~kW. Several  wavelengths  between 77
and 496~$\mu$m have been selected using NH$_3$, D$_2$O and
CH$_3$F as active media~\cite{bookGP}. The samples are irradiated under normal
incidence, along the growth direction. The terahertz radiation
causes indirect optical transitions within the lowest
size-quantized subband. Since the energy $\hbar \omega $ of
a terahertz photon is much smaller than the GaAs and InAs band gap as well as the
energy separation between the lowest lying filled subband and the
first excited subband, direct optical transitions are absent.
In experiments the angle between the polarization plane of the
linearly polarized light and the magnetic field is varied. This is
achieved by placing a  metal grid polarizer in a circularly
polarized beam obtained by  a crystalline quartz
$\lambda/4$-plate. Rotation of the metal grid enables us to vary,
at constant intensity, the angle $\alpha=0^\circ \div 180^\circ$
between the $x$ axis and the plane of linear polarization of the
light incident upon the sample (see inset in the upper panel of Fig.~3).
The external magnetic field with a maximum field strength of
$B = 0.6$~T 
is applied parallel to the heterostructure interface along [110] crystallographic direction. The
electric current generated by the light in unbiased devices is
measured via the voltage drop across a 50~$\Omega $ load resistor
in a closed circuit configuration. The voltage is recorded with a
storage oscilloscope. 

Irradiation of the samples at zero magnetic field does not lead --
as expected - to any current. A photocurrent response is obtained
only when the magnetic field is applied. The measured current pulses of 100~ns
duration reflect the corresponding laser pulses. As described by
Eqs.~(\ref{current1}) and~(\ref{spin}) the current increases
linearly with $B_y$ due to the increasing spin polarization (see 
upper panel of Fig.~4) and
changes sign upon reversal of $\bm{B}$. Corresponding data will be 
discussed below for the different samples.
The temperature and polarization dependences of the current
were measured in all samples
for two directions:  along and perpendicular  to the in-plane magnetic field. 
Figure~2 shows the
typical temperature dependence
of the photocurrent measured in the
direction perpendicular to the magnetic field and, as shown in the
inset, with the radiation also polarized perpendicularly to the magnetic
field. While the photocurrent is constant at low temperatures it
decreases as $1/T$ at temperatures above 100~K. As we show below
the peculiar temperature dependence is direct evidence that the
current is driven by the spin polarization given by
Eq.~(\ref{spin}). 

Before we discuss the corresponding microscopic origin in more
detail we present measurements of polarization dependences of
the current components perpendicular (Fig.~3, upper panel) and parallel
(Fig.~3, lower panel) to the applied magnetic field, $B_y$.
The polarization dependence of the current $j$ in
 transverse geometry   can be 
fitted by $j_x = j_1 \cos 2
\alpha + j_2$, and by $j_y = j_3 \sin 2 \alpha $  for the
longitudinal geometry. The overall polarization
dependences of the photocurrent remain  the same 
independently of temperature and
wavelength.
%
The increase of the radiation wavelength at
constant intensity results in an increased signal strength. The
wavelength dependence for both transverse and longitudinal configurations
is described by $j \propto \lambda^2$ in the
whole range of wavelengths used (see inset in
Fig.~3, lower panel) which corresponds to the spectral behaviour of  Drude
absorption, $\eta(\omega) \propto 1/\omega^2$ at $\omega \tau _p \gg 1$ (see~\cite{Seeger}). 
Here $\eta(\omega)$ 
is the  absorbance of heterostructure at the radiation
frequency $\omega$.

The fact that an offset $j_2$ is observed for the transverse geometry only is in accordance
with the phenomenological theory of magnetic field induced photocurrents~\cite{MPGE_jpcm}.
Under normal incidence of linearly polarized radiation
and in the presence of an in-plane magnetic field, $B_y$,
the current components are described by
\begin{equation}
\label{phen1} j_{x} =  C_1 B_{y} 
\left( e_{x}^2 - e_{y}^2 \right) I + C_2  B_{y}I\:,
\end{equation}
\begin{equation}
\label{phen2} j_{y} = C_3 B_{y} e_{x} e_{y} I\:,
\end{equation}
where $I$ and $\bm{e}$ are the light intensity and polarization vector, respectively.
The parameters $C_1$ to $C_3$ are coefficients determined by the
C$_{2v}$ symmetry  relevant for (001)-oriented structures. 
The polarization independent offset is described
by the second term in the right hand side of Eq.~(\ref{phen1})
and is present for the transverse geometry only. The only visible
consequence of this contribution is the offset in the upper panel of Fig.~3.
The other terms in
the right hand sides of Eqs.~(\ref{phen1}) and~(\ref{phen2})
yield polarization dependences in full agreement with
experiments.

All experimental features, i.e., the temperature and polarization dependences, are
driven by the spin degree of freedom: For fixed polarization for both excitation 
(Fig.~1~(a)) and relaxation (Fig.~1~(b))
mechanisms the current is proportional to the frequency dependent absorbance $\eta(\omega)$,
momentum relaxation time $\tau_p$, light intensity $I$ and 
average spin $S$: $j\propto \eta(\omega) I \tau_p S$.
Such type of expression, which determines the temperature dependence, 
is valid for fixed scattering mechanism, e.g. phonon or impurity scattering.
To corroborate this claim and obtain the polarization dependence microscopically
we present the  results of the corresponding theory for impurity scattering.

A
consistent description of  magneto-induced photocurrent can be
 developed within the framework of the spin-density
matrix. The scattering asymmetry induced contribution to the 
magneto-induced photocurrents is given by
\begin{equation}
\label{current} \bm{j} = \sum_{s\bm{k}} e \, \bm{v}_{\bm{k}} \,
\delta f_{s\bm{k}} = e \frac{2 \pi}{\hbar} \sum_{s\bm{k}
\bm{k}^\prime } \tau_p \, \left(\bm{v}_{\bm{k}} -
\bm{v}_{\bm{k}^\prime}\right) |M_{s \bm{k}, s \bm{k}^\prime}|^2 \,
\left(f_{s\bm{k}^\prime} - f_{s\bm{k}}\right) \, \delta(
\varepsilon_{\bm{k}} - \varepsilon_{\bm{k}^\prime} - \hbar \omega
) \:.
\end{equation}
Here $\bm{v}_{\bm{k}}=\hbar \bm{k}/ m^*$ is the electron velocity,
$m^*$ the effective electron mass, $\delta f_{s\bm{k}}$ the
fraction of the carrier distribution function stemming from
optical transitions in the spin subband~$s$,  $M_{s \bm{k}, s \bm{k}'}$ the matrix element of
the indirect optical transition, $f_{s\bm{k}}$ the equilibrium
distribution function, $\varepsilon_{\bm{k}}=\hbar^2 k^2/2m^*$ the
electron kinetic energy for in-plane motion, and $s$ an index
enumerating subbands with  spin states $\pm 1/2$ along the
direction of the external magnetic field.

To first order in spin-orbit interaction the compound matrix
element for the indirect optical transitions via impurity scattering has the
form~\cite{ac_field}
\begin{equation}\label{M}
M_{\bm{k}, \bm{k}^\prime} = \frac{eA}{c\,\omega m^*}\,
\bm{e}\cdot(\bm{k}-\bm{k}^\prime) V_{\bm{k} \bm{k}^\prime}
- 2\frac{eA}{c\hbar} \sum_{\alpha\beta}
V_{\alpha\beta}\,\sigma_{\alpha}e_{\beta} \:.
\end{equation}
Here $\bm{A} = A \bm{e}$ is the vector potential of the
electromagnetic wave, $c$ the light velocity and $V_{\bm{k}
\bm{k}^\prime}$ the scattering matrix element
given by~\cite{bulli}
\begin{equation}\label{scattering}
V_{\bm{k} \bm{k}^\prime} = V_0 + \sum_{\alpha\beta}
V_{\alpha\beta} \sigma_{\alpha} (k_{\beta} + k_{\beta}^\prime)\:,
\end{equation}
where the term $V_0$ describes the conventional spin independent scattering 
and the term proportional to the second rank pseudo-tensor $ V_{\alpha\beta}$ 
yields the asymmetric spin-dependent contribution linear in $\bm{k}$ and
 responsible for the effects described here. 
The first term on the right side of Eq.~(\ref{M}) describes transitions involving
virtual intermediate states in the conduction band while the
second term corresponds to transitions via virtual intermediate
states in the valence band. 


For C$_{2v}$ point-group symmetry
there are only two non-zero components of the tensor $ V_{\alpha\beta}$: 
$ V_{xy}$ and $ V_{yx}$.
By using Eqs.~(\ref{current}) to~(\ref{scattering}) 
an expression for the electric current $\bm{j}$ can be
derived. We consider the free-carrier absorption to be accompanied
by electron scattering from short-range static defects and assume
therefore that the matrix element $V_0$ and the coefficients
$ V_{\alpha\beta}$ are wavevector independent. As well as in experiment we consider linearly
polarized light at normal incidence and an
in-plane magnetic field $B_y$ resulting in an average spin $S_y$.
Then currents parallel and perpendicular to the magnetic field  can be written as
\begin{equation}\label{polariz1}
j_{x} = - 2(e^2_{x}-e^2_{y})\,  V_{yx} S_{y} \frac{e \tau_p}{\hbar
V_0} \,I \eta(\omega)\:,
\end{equation}
\begin{equation}\label{polariz2}
j_{y} =  - 4 e_{x} e_{y}  V_{xy} S_{y} \frac{e \tau_p}{\hbar V_0}
\,I \eta(\omega)\:,
\end{equation}
where  the photon energy $\hbar\omega$
is assumed to be smaller than the characteristic energy $\bar\varepsilon$.
Note, that the polarization independent part of Eq.~(\ref{phen1}) is missing here
as the above theory does not contain the relaxation
process, sketched in Fig.~1~(b), which is responsible for the background
signal of Fig.~3~(a).
%

Equations~(\ref{polariz1}) and~(\ref{polariz2}) contain the polarization dependence 
for transverse and longitudinal orientation, respectively, 
given by
\begin{equation}
    \label{trigon}
e^2_{x} - e^2_{y} = {\rm cos}\,2\alpha \,\,,\;\;\; 2 e_{x} e_{y} = {\rm
sin}\,2\alpha\,\,.
\end{equation}
%
The observed polarization dependences  are in a
full agreement with
Eqs.~(\ref{polariz1}), (\ref{polariz2}) and~(\ref{trigon}) (see
fits in Fig.~3). It should be noted that the 
polarization behaviour  of $j_x$ 
and $j_y$ depends neither on temperature nor on wavelength. It is 
solely described by  Eqs.~(\ref{phen1}) and~(\ref{phen2}) and does not
depend on specific scattering mechanism of Drude absorption.

However, the different scattering mechanisms involved in  Drude
absorption are reflected in the temperature dependence of the
photocurrent displayed in Fig.~2. 
While impurity scattering
prevails at low temperatures, phonon scattering takes over for $T
> 100$~K and is then the dominant scattering mechanism~\cite{kelly}. 
For
temperatures up to about 25~K the photocurrent is constant though
both, mobility and carrier density change significantly. 
Since, Drude absorption, $\eta
(\omega ) \propto n_s /\tau _p $ at $\omega \tau _p \gg 1$ (see~\cite{Seeger})
and at low temperatures $S\propto 1/\varepsilon _F \propto
1/n_s $ (see Eq.~(\ref{spin})), the current $j / I \propto \tau _p \,\eta (\omega )
S$ is constant and independent of $\tau _{p}$ and $n_{s}$. In
additional experiments we changed the carrier density at 4.2~K by
visible and near infrared light. 
For the sample 1, e.g.,  the carrier density (mobility) increases from 
$1.3\cdot 10^{11}$~cm$^{-2}$ ($1.7\cdot 10^6$~cm$^2$/Vs) to $3.0 \cdot 10^{11}$~cm$^{-2}$   
($4.1\cdot 10^6$~cm$^2$/Vs) after illumination at low $T$.
Though both $n_s$ and $\tau_p$  
increase by a factor of 2, the photocurrent remains unchanged, thus
confirming the above arguments.
In contrast, for $T >$ 100~K the carrier density $n_{s}$ is
roughly constant (saturation region) but $S$ is now $\propto 1/k_B T$,  
see
Eq.~(\ref{spin}).
Hence, the current $j$ is proportional to $n_s /T$
and becomes temperature dependent in accordance with our experimental
observation. Fits to the data in the low-$T$ and \mbox{high-$T$} regimes
are shown as solid lines in Fig.~2.
In the intermediate range of temperatures between 25 K and 100
K, where a rising current is observed, such simple analysis
fails. In this range the scattering mechanism, responsible for
the spin currents, changes from impurity dominating to phonon
dominating. This transition region is not yet theoretically treated
and out of scope of the present letter.

The experiments, carried out on different samples, are summarized 
in Fig.~4. 
Using the arrangement of upper panel in Fig.~3 for two fixed polarization directions, 
$\alpha = 0^\circ$ and $\alpha = 90^\circ$, we obtain
a linear increase of the corresponding 
photocurrent, shown in the upper panel of Fig.~4. 
By adding and subtracting the currents of both orientations the coefficients $j_1$ 
(polarization dependent amplitude) and $j_2$ 
(polarization independent background) can be extracted. 
Corresponding results of $j_1$ (lower left panel) and $j_2$ 
(lower right panel) for four different samples are shown. 
Due to the larger $g$-factor of sample~4 (InAs QW), 
causing larger average spin  $S$, the currents are
 largest for this sample. 
The other three samples are GaAs based heterostructures which differ 
in structural inversion asymmetry. Sample 1 is a heterojunction (see Table~I) 
which, due to triangular confinement potential, is expected to have the strongest SIA contribution. 
Samples 2 and 3 are quantum wells of the same width asymmetrically and symmetrically modulation 
doped, with larger and smaller strength of SIA,
respectively. 
The fact that with decreasing strength of the SIA coupling coefficient 
(from sample 1 to 3) the 
currents become smaller is in excellent agreement with our picture 
of asymmetric scattering  driven currents. 
The coupling strength constant controls 
the current via $V_{\alpha \beta}$ in Eqs.~(\ref{polariz1}), (\ref{polariz2}) and
equivalent expressions for other scattering mechanisms: The larger  the coupling strength 
the larger is the effect of asymmetric scattering.


Finally we would like to address the role of spin-dependent relaxation, 
sketched in Fig.~1~(b).
%
%
The  absorption of radiation
leads to an electron gas heating. Due to the spin-dependent asymmetry of
scattering energy relaxation rates for positive and negative $k_x$ within each spin subband
are nonequal as indicated by 
bent arrows of different thicknesses.
%
%
Thus, spin flows $i^\prime_{\pm 1/2}$  are
generated in both spin subbands.
Like for Drude excitation
spin separation takes place and applying
a magnetic field results in a net electric current. 
As indicated in Fig.~1~(a) and Fig.~1~(b) excitation and relaxation induced
currents flow in opposite directions. Experimentally this is observed for all samples: $j_1$
and $j_2$ for each sample have consistently opposite signs. As mentioned before $j_1$  stands 
 for polarization dependent part of the photocurrent due to excitation, while $j_2$
 is polarization independent and due to relaxation. 
%
%
%
%

Summarizing, we emphasize that all central  experimental features of the 
terahertz photocurrent, namely, magnetic field, temperature, mobility, 
and concentration dependences
provide evidence that the observed effect is solely determined by 
the spin degree of freedom. 
Furthermore our observations suggest that 
heating of the  electron gas by any means (microwaves, voltage  etc.) is sufficient 
to generate spin currents. Our results demonstrate that spin-dependent 
scattering provides a new tool for spin manipulation.

\subsection*{Acknowledgements} 
We would like to thank Imke Gronwald for preparation of the samples. 
This work
was supported by the DFG via Project GA-501/5, Research Unit
FOR370 and Collaborative Research Center SFB689, the RFBR and
programs of the RAS. S.G. thanks the HBS and S.A.T. the Foundation
``Dynasty'' - ICFPM and the President Grant for young scientists
for support.

\newpage

\begin{table}
\caption{Parameters of investigated samples. Mobility and electron sheet density data 
are obtained at 4.2~K in the dark.\\[1em]}

\begin{tabular}{ccccccc}

\hline  
\,\,\,\, sample  & \,\,\,\, material & \,\,\,\, QW width & \,\,\,\, spacer 1 & \,\,\,\, spacer 2 & \,\, \,\,mobility  &  \,\,\,\, density \\
        &          &     \AA\     &   \AA\        &   \AA\        & \,\,\,\,cm$^2$/Vs  & \,\,\,cm$^{-2}$ \\
\hline       
\#1     & GaAs/AlGaAs     & $\infty$ &  500 &           & $1.7 \cdot 10^6$   & $1.3 \cdot 10^{11}$  \\
\#2     & GaAs/AlGaAs     & 300      &  700 &           & $3.6 \cdot 10^6$    & $1.3 \cdot 10^{11}$ \\
\#3     & GaAs/AlGaAs     & 300      &  700 & 1000      & $3.4 \cdot 10^6$    & $1.8 \cdot 10^{11}$ \\
\#4     & InAs/AlGaSb     & 150      &      &           & $2.0 \cdot 10^4$   & $1.3 \cdot 10^{12}$\\
\hline  
\end{tabular}
\end{table}

\begin{figure}
\centerline{\epsfxsize 0.5\linewidth \epsfbox{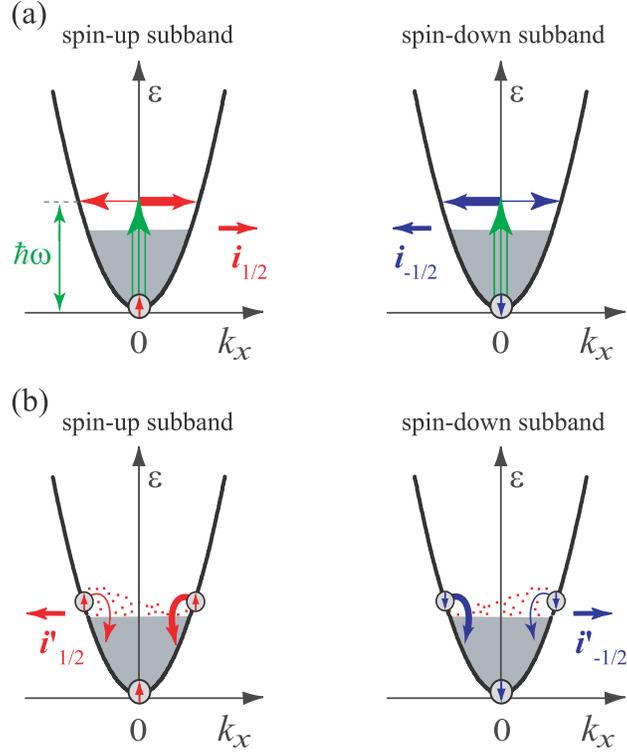}}
\caption{
 Microscopic origin of the zero-bias spin separation due to 
linear in  $\bm{k}$ and the Pauli spin matrices $\bm{\sigma}$ terms in 
electron  scattering, see~Eq.~(\ref{scattering}).
(a) caused by
excitation asymmetry and (b) due to spin-dependent energy relaxation.
It is assumed that scattering for  spin-up subband has larger probability 
for positive $k_x$ than that for negative $k_x$ and vice versa for  spin-down subband.
} 
\label{model2}
\end{figure}

\begin{figure}
\centerline{\epsfxsize 0.5\linewidth \epsfbox{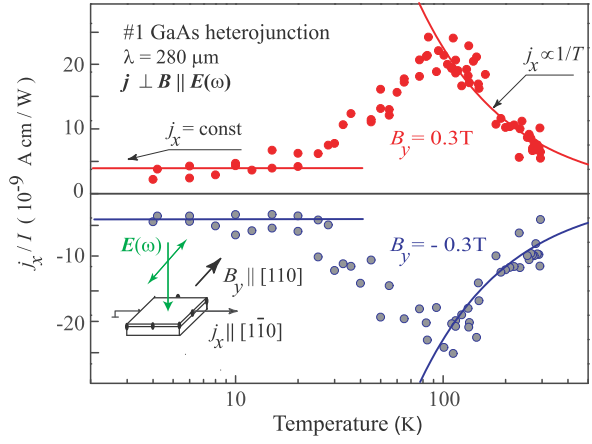}}
\caption{
 Photocurrent $j_x$ in a (001)-grown GaAs/AlGaAs heterojunction  as a function of
sample temperature for two opposite polarities of the magnetic field. The photocurrent $\bm j \, \bot \, \bm B
\,\|\,y$ is excited by normally incident linearly polarized
radiation of $\lambda = 280$~$\mu$m. Full lines are fits to $j_x = \pm \rm{const}$ at low $T$ and
$j_x \propto\pm 1/T$ at high $T$. The inset
shows the geometry of the experiment.
}
 \label{temperat}
\end{figure}

\begin{figure}
\centerline{\epsfxsize 0.5\linewidth \epsfbox{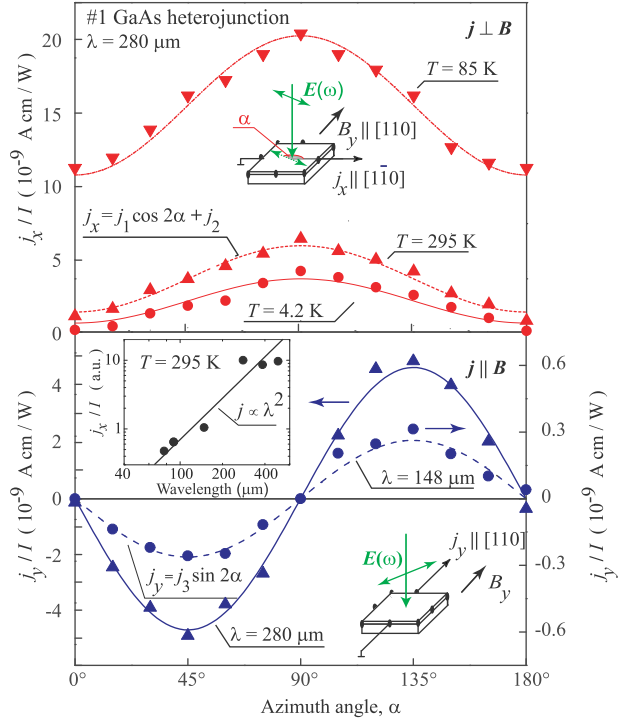}}
\caption{
 Photocurrents in a (001)-grown GaAs/AlGaAs heterojunction
as a function of the azimuth angle $\alpha$.
Data are obtained for normal incidence
of linearly polarized radiation 
measured for $B$~=~0.3~T. Upper panel: photocurrent $\bm j \, \bot
\, \bm B \,\|\,y$ at  $\lambda = 280$~$\mu$m and various
temperatures. Lines are fitted according to
$j_x= j_1 \cos 2\alpha + j_2$,  see Eqs. ~\protect(\ref{phen1}), ~\protect(\ref{polariz1}) 
and~\protect(\ref{trigon}).
Lower panel: photocurrent $\bm j \, \| \,
\bm B \,\|\,y$ measured at room temperature for two wavelengths
$\lambda =$ 148 and 280~$\mu$m. Lines are fitted according
to $j_y = j_3 \sin 2\alpha$, see Eqs. ~\protect(\ref{polariz2}) and~\protect(\ref{trigon}).
Insets show the experimental geometries. An additional inset in lower panel
demonstrates the wavelength dependence of the signal for the
transverse geometry. The full line shows $j_x \propto\lambda^2$.
}
 \label{fig3f}
\end{figure}

\begin{figure}
\centerline{\epsfxsize 0.5\linewidth \epsfbox{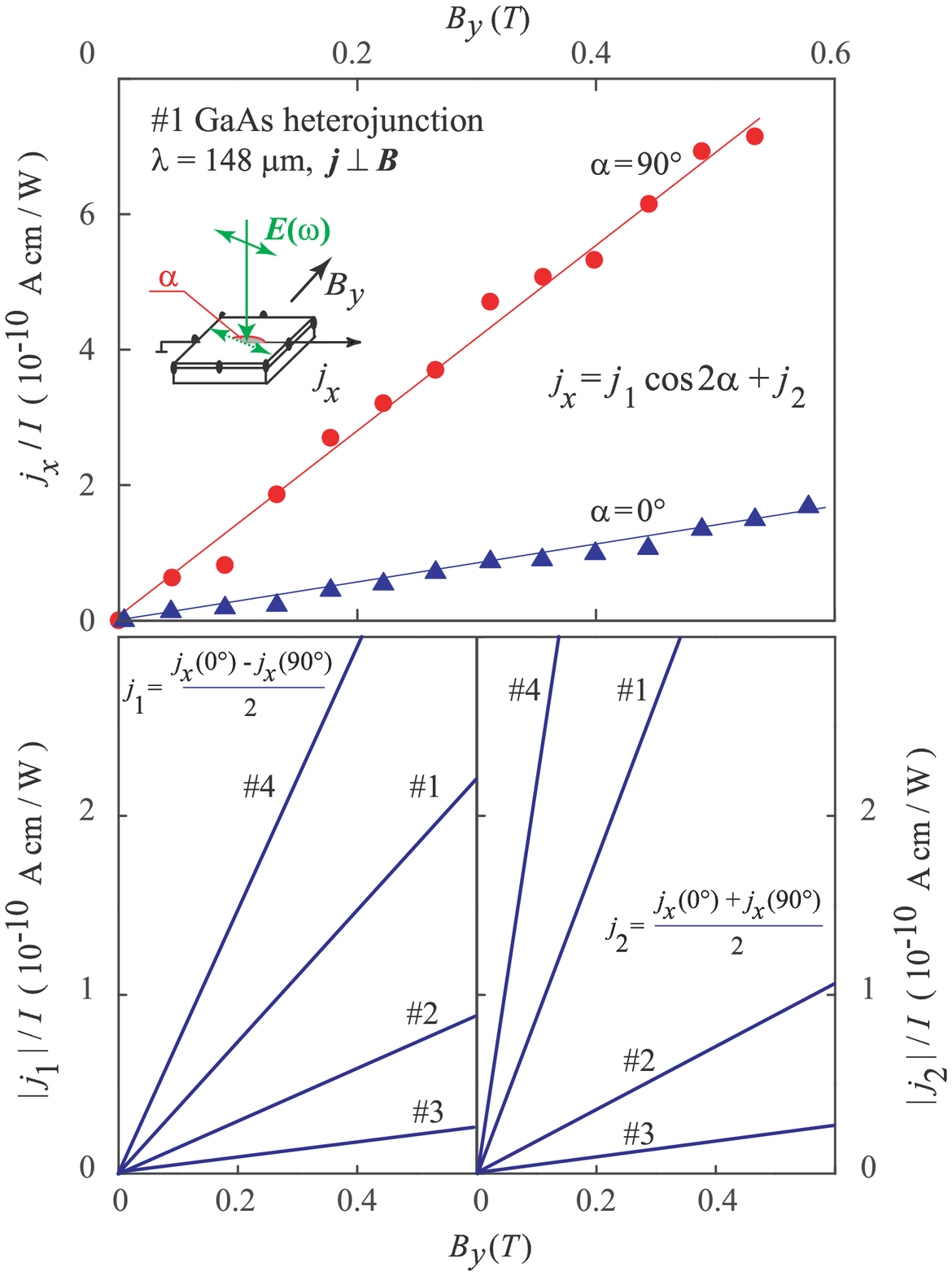}}
\caption{
 Magnetic field dependence of the transversal photocurrent.
Upper panel: $j_x(B)$ measured in sample~1 at room temperature for two polarization states.
Lower panels: $j_1(B)$ (left panel) and $j_2(B)$ (right panel) obtained by adding and subtracting the currents 
corresponding to excitations of both polarizations for various samples.
}
 \label{fig4f}
\end{figure}


\begin{thebibliography}{99}
\bibitem{spinHall} Spin-dependent scattering 
processes responsible for the zero-bias spin separation and for the extrinsic spin Hall effect, 
in spite of the vocabulary similarity, are  principally different. While 
odd in $\bm k$ scattering, does not require 
directed electron motion and needs sufficiently low symmetry of material,
skew scattering
driving extrinsic spin Hall effect is due to average carrier velocity 
and can be  present even in highly symmetric materials.
\bibitem{footnote4} In fact all terms odd in ${\bm k}$, including ${\bm k}$-cubic terms, may also contribute to  spin dependent asymmetric scattering.
In the present paper we limit our consideration to the lowest order term which is linear in ${\bm k}$.
\bibitem{footnote1} Note that we assumed $k_x$~=~0, and structural inversion asymmetry only.
\bibitem{footnote2} Here we neglect effects of electron-electron
interaction on a spin transport recently addressed in $^{12}$.
\end{thebibliography}
\end{document}